\documentclass[aps,pra,superscriptaddress,amsmath,amssymb,preprintnumbers,showpacs,twocolumn]{revtex4-1}

\usepackage{amssymb}
\usepackage{bm}
\usepackage{color}
\usepackage{subfigure}
\usepackage[utf8x]{inputenc}
\usepackage[english]{babel}
\usepackage[T1]{fontenc}
\usepackage{lmodern}
\usepackage{bbold}
\usepackage{amsmath}
\usepackage{amsthm}
\usepackage[pdftex]{graphicx}
\usepackage{epstopdf}
\usepackage{pgfplots}

\begin{document}

\title{Nonlocal quantum memory effects in a correlated multimode field}

\author{Steffen Wi\ss mann}
\author{Heinz-Peter Breuer}
\affiliation{Physikalisches Institut, Universit\"{a}t Freiburg, Hermann-Herder-Stra\ss e 3, D-79104 Freiburg, Germany}

\begin{abstract}
We review the model of two qubits coupled locally to an environment
which consists of nonlocally correlated field modes 
[Phys.~Rev.Lett. \textbf{108}, 210402 (2012)]. We derive 
the correct expressions for the reduced dynamics of the two-qubit system 
and demonstrate that strong nonlocal memory effects are indeed present for 
suitable initial EPR-type Gaussian environmental states.
\end{abstract}

\pacs{03.65.Yz, 03.65.Ta, 03.67.Pp}

\maketitle

\section{Introduction}\label{sec:intro}
The study of non-Markovian effects in the dynamics of open quantum systems has 
attracted vast attention in recent years. Several suggestions for the quantification 
of non-Markovian behavior \cite{Wolf,Measure,MeasurePaper,Fisher,RHP} have 
been made, applied to different physical models \cite{Mazzola,Govinda} and 
compared among each others \cite{Chruscinski,Haikka,clMarkovQ}. Moreover, 
several experiments \cite{Nature,SingleQubitExp} have been performed 
quantifying non-Markovian behavior in terms of the flow of information between 
the open system and its environment \cite{Measure}. 

Recently, it has been shown theoretically as well as experimentally that
quantum memory effects can also be induced by nonlocal environmental
correlations \cite{NonlocalTheo,Nonlocal}. The first model studied in 
\cite{NonlocalTheo} illustrates this effect by means of an open two-qubit system 
coupled locally to an environmental multimode field in a nonlocally correlated
initial state. The coherence factors and the assumption on the initial correlated 
environmental states used in Ref.~\cite{NonlocalTheo} are however not correct. 
Here, we provide the correct expressions for the quantum dynamical map and the 
conditions on two-mode Gaussian states and show that strong nonlocal memory 
effects indeed occur for particular two-mode Gaussian states whose covariance 
matrix is in standard form. 

\section{Physical model}\label{sec:model}
The first model studied in Ref.~\cite{NonlocalTheo} regarding nonlocal memory 
effects consists of two qubits coupled to a bosonic environment.
Each of the two qubits interacts locally with its own multimode bosonic bath which 
is assumed to be a part of a correlated environment. For the state 
of the latter one chooses a product of two-mode Gaussian states correlating each 
pair of modes of the two bosonic baths.

The Hamiltonian of the total system is given by
\begin{equation}
 H=\sum_{i=1}^2(H_S^i+H_E^i+H_{int}^i)~,
\end{equation}
where $H_S^i=\epsilon_i\hat{\sigma}_z^i$ and $H_E^i=\sum_k\omega_k^i {\hat{b}_k}^{i\dag} \hat{b}_k^i$ with ${\hat{b}_k}^{i(\dag)}$ referring to the annihilation (creation) operator of the $k$th mode of bath $i$. The interaction Hamiltonian is build up by local interactions which obey
\begin{equation}\label{eq:Hint}
 H_{int}^i=\chi_i(t)\sum_k \hat{\sigma}_z^i\otimes(g_k^i {\hat{b}_k}^{i\dag} + {g_k^i}^* \hat{b}_k^i)~,
\end{equation}
where $g_k^i$ denotes the coupling strength of the $i$th subsystem. Without loss of generality we assume that the coupling strengths are real-valued, i.e. $g_k^i\in\mathbb{R}$ for $i=1,2$ and all $k$. The function $\chi_i(t)$ is given by
\begin{equation}
\chi_i(t)=\Theta(t-t_i^s)\Theta(t_i^f-t)=\begin{cases}
                                          1~ ,~ t\in[t_i^s,t_i^f]\\ 0~ ,~ \text{else}
                                         \end{cases}~,
 \end{equation}
for some $t_i^f>t_i^s>0$. It simulates the turning-on and -off of the local interactions of subsystem $i$ at time $t_i^s$ and $t_i^f$, respectively. The duration of the local interactions and the inset can be varied independently for both subsystems so that it is possible to switch from simultaneous to a successive application of the interactions. Without loss of generality we may assume $t_1^s\leq t_2^s$.

The dynamics of the model is conveniently solved in the interaction picture. Turning to this picture the interaction Hamiltonian $H_{int}$ transforms into $H_{int}^I(t)=\exp(+iH_0 t) H_{int}(t) \exp(-iH_0 t)$ where $H_0=\sum_i (H_S^i+H_E^i)$ which yields
\begin{equation}
H_{int}^I(t)=\sum_{j}\chi_j(t)\hat{\sigma}_z^j \otimes \sum_k (g_k^j e^{i\omega_k^j t}{\hat{b}_k}^{j\dag} + {g_k^j}^* e^{-i\omega_k^j t} \hat{b}_k^j)~,
\end{equation}
as $[\hat{b}_k^i,{\hat{b}_l}^{j\dag}]=\delta_{kl}\delta_{ij}$. Applying again this commutation relation for the annihilation and creation operators it can be shown that the interaction Hamiltonian $H_{int}^I(t)$ at times $t$ and $t'$ obeys
\begin{align}
 &[H_{int}^I(t),H_{int}^I(t')]=-2i \phi(t-t')~,
\end{align}
where $\phi(t-t')=\sum_{j,k}\chi_j(t)\chi_j(t') |g_k^j|^2 \sin[\omega_k^j(t-t')]$ is a scalar function. It is well known \cite{BreuerBuch} that the time evolution operator in the interaction picture is then given by
\begin{align}
 U_I(t)&=\mathrm{T}_\leftarrow \exp[-i\int_0^t \mathrm{ds} H_{int}^I(s)] \nonumber\\
 &=\exp[i\int_0^t \mathrm{ds}\int_0^t \mathrm{ds'}\phi(s-s')\Theta(s-s')]\nonumber\\
&\qquad\cdot \exp[-i\int_0^t \mathrm{ds} H_{int}^I(s)]~.
\end{align}
The time evolution operator thus consists of a phase factor $d(t)\equiv\exp[i\int_0^t \mathrm{ds}\int_0^t \mathrm{ds'}\phi(s-s')\Theta(s-s')]$ and a non-trivial operator $V(t)\equiv\exp[-i\int_0^t \mathrm{ds} H_{int}^I(s)]$ which can be rewritten as
\begin{align}
V(t)&=\exp[\sum_{j,k}\hat{\sigma}_z^j \otimes (\beta_k^j(t){\hat{b}_k}^{j\dag} - \beta_k^j(t)^*\hat{b}_k^j)]~,
\end{align}
with
\begin{equation}
 \beta_k^j(t)=\frac{g_k^j}{\omega_k^j}\,e^{i\omega_k^jt_j^s}\Bigl(1-e^{i\omega_k^j\int_0^t \mathrm{ds} \chi_j(s)}\Bigr)~,
\end{equation}
since $\int_o^t\mathrm{ds}\chi_j(s)e^{i\omega_k^j s}=i e^{i\omega_k^jt_j^s}(1-e^{i\omega_k^j\int_0^t \mathrm{ds} \chi_j(s)})/\omega_k^j$. Note that the phase factor $e^{i\omega_k^jt_j^s}$, taking into account the influence of the free evolution prior the inset of the interaction, is missing in Ref. \cite{NonlocalTheo}. Hence, $V(t)$ is a two-mode displacement or Weyl operator. The local unitaries $V_j(t)$ ($V(t)\equiv V_1(t)V_2(t)$) act therefore according to
\begin{align}
V_j(t)|0\rangle\otimes|\eta\rangle&=|0\rangle\otimes\prod_k D(-\beta_k^j(t))|\eta\rangle~, \\
V_j(t)|1\rangle\otimes|\eta\rangle&=|1\rangle\otimes\prod_k D(+\beta_k^j(t))|\eta\rangle~,
\end{align}
where $D$ denotes the displacement operator and $|0\rangle$, $|1\rangle$ refer to the ground and excited state of the two-level system, respectively. Moreover, $|\eta\rangle$ is an arbitrary pure state of the environment. Finally, the time evolution operator in the Schrödinger picture is given by
\begin{equation}
 U(t)=e^{-iH_0 t}U_I(t)~.
\end{equation}
Thus, the time evolution of the initially factorizing state
\begin{align}
 |\psi(0)\rangle&=|\psi_{12}\rangle\otimes|\eta_{12}\rangle~, \\
\intertext{where}
|\psi_{12}\rangle&=a_{00}|00\rangle+a_{01}|01\rangle+a_{10}|10\rangle+a_{11}|11\rangle~, \\
|\eta_{12}\rangle&=\bigotimes_k|\eta_{12}^k\rangle~,\label{eq:eta}
\end{align}
with $|\eta_{12}^k\rangle$ referring to arbitrary two-mode states of the $k$th mode of bath $1$ and $2$, reads
\begin{equation}
 |\psi(t)\rangle=e^{-iH_0 t}d(t)\sum_{n,m=0}^1 a_{nm}|nm\rangle\otimes|\eta_{12}^{nm}(t)\rangle~.
\end{equation}
Here, the time-evolved environmental states are given by $|\eta_{12}^{nm}(t)\rangle\equiv\bigotimes_k D((-1)^{n+1}\beta_k^1(t))\otimes D((-1)^{m+1}\beta_k^2(t))|\eta_{12}\rangle$.
The reduced state of the two two-level systems is then obtained by taking the partial trace over the environmental degrees of freedom which yields
\begin{align}\label{eq:genstate}
&\rho_S^{12}(t)\nonumber\\
=&\sum_{n,m,r,s=0}^1 e^{it\bigl\{[(-1)^n-(-1)^{r}]\epsilon_1+[(-1)^m-(-1)^{s}]\epsilon_2\bigr\}}\cdot a_{nm}a_{rs}^* \nonumber\\
&\qquad\qquad \cdot\langle \eta_{12}^{nm}(t)|\eta_{12}^{rs}(t)\rangle \cdot|nm\rangle\langle rs|\nonumber\\
=&\left(\begin{matrix}|a_{11}|^2 & a_{11}a_{10}^*\tilde{\kappa}_2(t) & a_{11}a_{01}^*\tilde{\kappa}_1(t) & a_{11}a_{00}^*\kappa_{12}(t) \\
 & |a_{10}|^2 & a_{10}a_{01}^*\Lambda_{12}(t) & a_{10}a_{00}^*\kappa_1(t) \\
 & &|a_{01}|^2 & a_{01}a_{00}^*\kappa_2(t)\\
 \text{c.c.}& & &|a_{00}|^2 \\ \end{matrix}\right)~,
\end{align}
where
\begin{align}\label{eq:coh}
 \kappa_1(t)&=e^{-2i\epsilon_1t}\langle\eta_{12}^{10}(t)|\eta_{12}^{00}(t)\rangle~, \\
 \kappa_2(t)&=e^{-2i\epsilon_2t}\langle\eta_{12}^{01}(t)|\eta_{12}^{00}(t)\rangle~, \\
\tilde{\kappa}_1(t)&=e^{-2i\epsilon_1t}\langle\eta_{12}^{11}(t)|\eta_{12}^{01}(t)\rangle~, \\
\tilde{\kappa}_2(t)&=e^{-2i\epsilon_2t}\langle\eta_{12}^{11}(t)|\eta_{12}^{10}(t)\rangle~, \\
 \kappa_{12}(t)&=e^{-2i(\epsilon_1+\epsilon_2)t}\langle\eta_{12}^{11}(t)|\eta_{12}^{00}(t)\rangle~, \\
 \Lambda_{12}(t)&=e^{-2i(\epsilon_1-\epsilon_2)t}\langle\eta_{12}^{10}(t)|\eta_{12}^{01}(t)\rangle~, \label{eq:cohl}
\end{align}
and 
\begin{align}
 &\langle\eta_{12}^{nm}(t)|\eta_{12}^{rs}(t)\rangle\nonumber\\
&=\prod_k\langle\eta_{12}^k|\left[D\bigl((-1)^{n+1}\beta_k^1(t)\bigr)\otimes D\bigl((-1)^{m+1}\beta_k^2(t)\bigr)\right]^\dag\nonumber\\
&\hspace{1.5cm}\left[D\bigl((-1)^{r+1}\beta_k^1(t)\bigr)\otimes D\bigl((-1)^{s+1}\beta_k^2(t)\bigr)\right]|\eta_{12}^k\rangle\nonumber\\
&\equiv\prod_k \chi_k^{nmrs}
\end{align}
Using the identities $D(\alpha)^\dag=D(-\alpha)$ and $D(\alpha)D(\beta)=e^{-2\textrm{Im}(\alpha^*\beta)}D(\alpha+\beta)$ for displacement operators one obtains for $\chi_k^{nmrs}$:
\begin{align}\label{eq:twomode}
&\chi_k^{nmrs}\\
&=\langle\eta_{12}^k|\exp\left[\sum_{j=1}^2\gamma_{k,nmrs}^j(t) {b_k^j}^\dag-\gamma_{k,nmrs}^j(t)^* b_k^j\right]|\eta_{12}^k\rangle~,\nonumber
\end{align}
with
\begin{align}
\gamma_{k,nmrs}^1(t)\equiv\{(-1)^n-(-1)^r\}\beta_k^1(t)~,\\
\gamma_{k,nmrs}^2(t)\equiv\{(-1)^m-(-1)^s\}\beta_k^2(t)~.
\end{align}
Hence, $\chi_k^{nmrs}$ is the Wigner characteristic function of the pure state $|\eta_{12}^k\rangle$ which is easily determined for two-mode Gaussian states.

\section{Coherence factors for two-mode Gaussian states}\label{sec:2modeGauss}
In the following we state the explicit expressions for the coherence factors \eqref{eq:coh}$-$\eqref{eq:cohl} if the environmental state $|\eta_{12}^k\rangle$ is chosen to be a two-mode Gaussian state whose covariance matrix is in standard form. This choice corresponds to the states considered in Ref. \cite{NonlocalTheo}. Without loss of generality one may assume that the Gaussian state has zero mean as this can always be achieved applying local operations \cite{GaussianIllumi,GaussianIndia}. This does not change the correlations in the two-mode state we are mainly interested in. 

We recall that a state of a continuous variable system 
$\rho\in\mathcal{S}(L^2(\mathbb{R}^n))$ is an $n$-mode Gaussian if and only if for all $\vec{x},\vec{y}\in\mathbb{R}^n$ the observable $\hat{Y}\equiv\sum_{j=1}^n(x_j\hat{p}_j-y_j\hat{q}_j)$ has a normal distribution on $\mathbb{R}^n$ in the state $\rho$ \cite{GaussianIndia} where 
\begin{equation}
 \hat{q}_j=\frac{1}{\sqrt{2}}(\hat{b}_j+\hat{b}_j^\dag)~~,~~\hat{p}_j=\frac{-i}{\sqrt{2}}(\hat{b}_j-\hat{b}_j^\dag)~,
\end{equation}
define the canonical position and momentum operators. That is, one has
\begin{align}\label{eq:chiGen}
 \chi_{Y,\rho}^t(\vec{z})&=\mathrm{Tr}(\rho \exp[-i t \hat{Y}])\nonumber\\
 &=\exp\left[-it(\boldsymbol{l}^T\vec{x}-\boldsymbol{m}^T\vec{y})-\frac{t^2}{2}\vec{w}^T\boldsymbol{S}\vec{w}\right]~,
\end{align}
where $\vec{w}^T=(y_1,x_1,\dots,y_n,x_n)$ and $\boldsymbol{l}_i=\langle\hat{p}_i\rangle$, $\boldsymbol{m}_i=\langle\hat{q}_i\rangle$ denote the mean position and momentum. Moreover, the $2n\times 2n$-matrix $\boldsymbol{S}$ is the covariance matrix of the operator $\hat{X}'=(\hat{q}_1,-\hat{p}_1,\dots,\hat{q}_n,-\hat{p}_n)$, i.e.
\begin{align}
\boldsymbol{S}&=\boldsymbol{\sigma}_{X'}\equiv \left(\Bigl(\tfrac{1}{2}\langle\{\hat{X}'_j,\hat{X}'_j\}\rangle-\langle\hat{X}'_j\rangle\langle\hat{X}'_j\rangle\Bigr)_{ij}\right)~.\label{eq:covS}
\end{align}
For $t=\sqrt{2}$ the operator $\exp[-i t \hat{Y}]$ is the Weyl operator $\mathcal{W}(z)$
\begin{align}\label{eq:Weyl}
 \mathcal{W}(z)&=\exp\left[\sum_{j=1}^n(z_j\hat{b}_j^\dag-z_j^*\hat{b}_j)\right]~,
\end{align}
where $z_j=x_j+i y_j$ for all $j$. One can show \cite{GaussianIndia,SimonGaussian} that the right hand side of Eq. \eqref{eq:chiGen} defines the characteristic function of an $n$-mode Gaussian state for some $\boldsymbol{l},\boldsymbol{m}\in\mathbb{R}^n$ and $\boldsymbol{S}\geq0$ if and only if
\begin{align}\label{eq:conCova}
\boldsymbol{S}+\frac{i}{2}\Omega_n\geq0~,
\end{align}
where the symplectic form $\Omega_n=\oplus_{k=1}^n \omega$ with $\omega=\left(\begin{matrix}0&1\\-1&0\end{matrix}\right)$ encodes the canonical commutation relations. This condition is sometimes called the Robertson-Schrödinger uncertainty relation and is a direct consequence of the Schrödinger uncertainty relation and Williamson's theorem \cite{Williamson} which states that any real-valued, symmetric and positive matrix can be transformed into a diagonal form by an appropriate symplectic operation \cite{SimonGaussian}. Note that Eq. \eqref{eq:conCova} implies positivity of $\boldsymbol{S}$ and that one has $\boldsymbol{S}+(i/2)\Omega_n\geq0$ if and only if $\boldsymbol{S}-(i/2)\Omega_n\geq0$. 

Now, suppose the environmental states $|\eta_{12}^k\rangle$ of Eq.~\eqref{eq:eta} are a two-mode Gaussian state with zero mean. According to \eqref{eq:chiGen}, Eq. \eqref{eq:twomode} is then given by
\\
\begin{align}\label{eq:char1}
 &\chi_k^{nmrs}\bigl((\gamma_{k,nmrs}^1(t),\gamma_{k,nmrs}^2(t)\bigr)\nonumber\\
 &=\exp\left[-\vec{\lambda}_{k,nmrs}(t)^T\boldsymbol{S}_k\vec{\lambda}_{k,nmrs}(t)\right]~,
\end{align}
where 
\begin{align}
\vec{\lambda}_{k,nmrs}(t)\equiv\left(\begin{matrix}\textrm{Im}(\gamma_{k,nmrs}^1(t))\\\textrm{Re}(\gamma_{k,nmrs}^1(t))\\\textrm{Im}(\gamma_{k,nmrs}^2(t))\\\textrm{Re}(\gamma_{k,nmrs}^2(t))\end{matrix}\right)~,
\end{align}
and $\boldsymbol{S}$ satisfies \eqref{eq:conCova}. We point out that the covariance matrix which is considered in Ref. \cite{NonlocalTheo} violates Eq. \eqref{eq:conCova} for any $c\neq0$. 

For any two-mode covariance matrix $\boldsymbol{S}$ there exist local symplectic operations such that the expectation values $\langle\{\hat{q}_i,\hat{p}_j\}\rangle$ are removed \cite{GaussianZoller,GaussianSimon} so that $\boldsymbol{S}$ is transformed into the so called standard form
\begin{align}
\boldsymbol{S}_{\mathrm{sf}}&\equiv\left(\begin{matrix} a&0&c_+&0\\0&a&0&c_-\\c_+&0&b&0\\0&c_-&0&b                                                                                                                                                                                                                                                                                                                       \end{matrix}\right)~,\label{eq:standardform1}
\end{align}
where $a,b,c_\pm\in\mathbb{R}$ and $a,b\geq1/2$. 

The coherence factors defined in Eq. \eqref{eq:coh}-\eqref{eq:cohl} can now be written as products of characteristic functions, i.e. 
\begin{align}
 \kappa_1(t)&=e^{-2i\epsilon_1t}\nonumber\\
 &\qquad\cdot\exp\left[-\sum_k \vec{\lambda}_{k,1000}(t)^T \boldsymbol{S}_{k} \vec{\lambda}_{k,1000}(t)\right]~,\label{eq:kappa1P} \\
 \kappa_2(t)&=e^{-2i\epsilon_2t}\nonumber\\
 &\qquad\cdot \exp\left[-\sum_k \vec{\lambda}_{k,0100}(t)^T \boldsymbol{S}_{k} \vec{\lambda}_{k,0100}(t)\right]~,\label{eq:kappa2P} \\
\kappa_{12}(t)&=e^{-2i(\epsilon_1+\epsilon_2)t}\nonumber\\
&\qquad\cdot \exp\left[-\sum_k \vec{\lambda}_{k,1100}(t)^T \boldsymbol{S}_{k} \vec{\lambda}_{k,1100}(t)\right]~,\label{eq:kappa12P} \\
 \Lambda_{12}(t)&=e^{-2i(\epsilon_1-\epsilon_2)t}\nonumber\\
 &\qquad \cdot \exp\left[-\sum_k\vec{\lambda}_{k,1001}(t)^T \boldsymbol{S}_{k} \vec{\lambda}_{k,1001}(t)\right]~,\label{eq:lambda12P} 
\end{align}
and $\tilde{\kappa}_j(t)=\kappa_j(t)$. Henceforth, we assume that the Gaussian states are identical for all modes, i.e. $\boldsymbol{S}=\boldsymbol{S}_k$ for all $k$. For a general covariance matrix in standard form, the exponentials in 
Eqs.~\eqref{eq:kappa1P}-\eqref{eq:lambda12P} can be evaluated employing Laplace transforms. After performing the continuum limit for an ohmic spectral density $J_j=\alpha_j \omega \exp[-\omega/\omega_c]$ with equal cutoff frequency $\omega_c$ but different couplings $\alpha_j$ for the two bosonic baths, one obtains expressions containing the Laplace transform of $(1-\cos(y t))/t$ and $\sin$-modulated functions in the exponentials which can be evaluated using standard techniques. For a covariance matrix in standard form with real-valued coefficients $a$, $b$ and $c_\pm$ one then obtains for the coherence factors \eqref{eq:kappa1P}-\eqref{eq:lambda12P}:
\begin{widetext}
 \begin{align}
  \kappa_1(t)&=e^{-2i\epsilon_1t}\left(1+\omega_c^2t_1(t)^2\right)^{-4 a \alpha_1}~,\label{eq:kappa1}\\
  \kappa_2(t)&=e^{-2i\epsilon_2t}\left(1+\omega_c^2t_2(t)^2\right)^{-4 b \alpha_2}~,\label{eq:kappa2}
\end{align}
\begin{align}
  \kappa_{12}(t)&=\frac{e^{-2i(\epsilon_1+\epsilon_2)t}}{(1+\omega_c^2t_1(t)^2)^{4a\alpha_1}(1+\omega_c^2t_2(t)^2)^{4b\alpha_2}}\left(\frac{(1+\omega_c^2{t_2^s}^2)(1+\omega_c^2(t_1(t)-t_2(t)-t_2^s)^2)}{(1+\omega_c^2(t_1(t)-t_2^s)^2)(1+\omega_c^2(t_2(t)+t_2^s)^2)}\right)^{4c_-\sqrt{\alpha_1\alpha_2}}\nonumber\\
&\qquad\cdot\left(\frac{\bigl(1+\omega_c^{2}(t_1(t)-t_2^s)^2\bigr)\bigl(1+\omega_c^{2}(t_2(t)+t_2^s+t_1(t))^2\bigr)}{\bigl(1+\omega_c^{2}(t_1(t)+t_2^s)^2\bigr)\bigl(1+\omega_c^{2}(t_2(t)+t_2^s-t_1(t))^2\bigr)}\right)^{2(c_--c_+)\sqrt{\alpha_1\alpha_2}}~,\label{eq:kappa12}\\
\Lambda_{12}(t)&=\frac{e^{-2i(\epsilon_1-\epsilon_2)t}}{(1+\omega_c^2t_1(t)^2)^{4a\alpha_1}(1+\omega_c^2t_2(t)^2)^{4b\alpha_2}}\left(\frac{(1+\omega_c^2{t_2^s}^2)(1+\omega_c^2(t_1(t)-t_2(t)-t_2^s)^2)}{(1+\omega_c^2(t_1(t)-t_2^s)^2)(1+\omega_c^2(t_2(t)+t_2^s)^2)}\right)^{-4c_-\sqrt{\alpha_1\alpha_2}}\nonumber\\
&\qquad\cdot\left(\frac{\bigl(1+\omega_c^{2}(t_1(t)-t_2^s)^2\bigr)\bigl(1+\omega_c^{2}(t_2(t)+t_2^s+t_1(t))^2\bigr)}{\bigl(1+\omega_c^{2}(t_1(t)+t_2^s)^2\bigr)\bigl(1+\omega_c^{2}(t_2(t)+t_2^s-t_1(t))^2\bigr)}\right)^{2(c_+-c_-)\sqrt{\alpha_1\alpha_2}}~,\label{eq:lambda12}
\end{align}
\end{widetext}
where we have set $t_1^s=0$ for simplicity. The time $t_2^s$, at which the interaction of the second spin with its bath is turned on, remains however arbitrary. 

\section{EPR-type initial state}
A particular candidate for a two-mode Gaussian state whose covariance matrix $\boldsymbol{\sigma}_X$ for $\hat{X}=(\hat{q}_1,\hat{p}_1,\hat{q}_2,\hat{p}_2)$ is in standard form \eqref{eq:standardform1} is given by the EPR-type state \cite{Werner}
\begin{equation}\label{eq:WernerEnergy}
 |\psi_u\rangle=\sqrt{1-u^2}\sum_{n=0}^\infty u^n |n\rangle\otimes|n\rangle~,
\end{equation}
where $u=\tanh(r)$. This state is the analog of a maximally entangled state for continuous variable systems as it corresponds to the Schmidt-decomposition of a maximally entangled state for $r\rightarrow\infty$. The state represents the physical realization of the model used by Einstein, Podolski and Rosen in their famous Gedankenexperiment \cite{EPR}. The variable $r\in\mathbb{R}$ denotes the squeezing parameter and is a reminiscence that this state is obtained by squeezing the two-mode vacuum. Values about $r=5$ can be realized in experiments \cite{Werner}. 

In the position representation the wave function takes the form
\begin{eqnarray}\label{eq:WernerWave}
 \lefteqn{ \psi_u(q_1,q_2) = } \\ 
 && \frac{1}{\sqrt{\pi}} 
 \exp\left[-\frac{1}{4}\frac{1-u}{1+u}(q_1+q_2)^2
 -\frac{1}{4}\frac{1+u}{1-u}(q_1-q_2)^2\right]. \nonumber
\end{eqnarray}
We see that the exponent is dominated by the second term in the limit 
$r\rightarrow +\infty$ ($u\rightarrow +1$), yielding a wave function with strong 
positive correlations of the positions of the particles. In the opposite limit 
$r\rightarrow -\infty$ ($u\rightarrow −1$) the wave function describes strong 
anti-correlations between the particle positions.

The EPR-type state is also referred to as twin-beam or squeezed vacuum state and defines a two-mode Gaussian state with zero means as one can show by a direct calculation of the characteristic function $\chi_{Y,\rho}^t$ \eqref{eq:chiGen}. Its covariance matrix for $\hat{X}$, which can be also easily derived, is given by 
\begin{equation}\label{eq:WernerCov}
 \boldsymbol{\sigma}_{X,r}^\mathrm{EPR}\equiv\frac{1}{2}\left(\begin{smallmatrix} \cosh(2r)&0&\sinh(2r)&0\\0& \cosh(2r)&0&-\sinh(2r)\\\sinh(2r)&0& \cosh(2r)&0\\0&-\sinh(2r)&0& \cosh(2r)                                                                                                                                                                                                                                                                                                                       \end{smallmatrix}\right)~.
\end{equation}
We remark that according to our conventions the expressions for the wave 
function \eqref{eq:WernerWave} and the covariance matrix \eqref{eq:WernerCov} 
differ from what one typically finds in the literature (see, e.g., \cite{Werner}). The 
commonly stated covariance matrix of the twin-beam state lacks an overall factor 
of $1/2$ and a factor of $2$ in the 
arguments of the $\sinh$- and $\cosh$-terms. We also note that within the
conventions used in Ref.~\cite{NonlocalTheo} the correct covariance matrix is
obtained from \eqref{eq:WernerCov} by omitting the overall factor of $1/2$
and by reversing the signs of the $\sinh$-terms.

\section{Maximal backflow of information}\label{sec:backflow}
An established measure for the degree of non-Markovianity of the dynamics of an open quantum system is given by \cite{Measure,OptimalPair}
\begin{align}\label{eq:measure}
 \mathcal{N}(\Phi)\equiv\max_{\rho_1\perp\rho_2}\int_{\sigma>0}\mathrm{d}t~ \sigma(t,\rho_1,\rho_2)~,
\end{align}
where $\rho_1$, $\rho_2$ are two orthogonal states of the open system and
\begin{align}
 \sigma(t,\rho_1,\rho_2)\equiv\frac{\mathrm{d}}{\mathrm{d}t}\mathcal{D}\bigl(\Phi_t(\rho_1),\Phi_t(\rho_2)\bigr)
\end{align}
describes the dynamical change of the trace distance $\mathcal{D}$ of these states. Moreover, the set $\Phi=\{\Phi_t|0\leq t\leq T\}$ denotes the one-parameter family of dynamical mappings which describe the dynamics of the open system. Hence, the measure $\mathcal{N}$ determines the maximal increase of the trace distance for any pair of orthogonal input states.

Employing this tool to quantify memory effects in our pure dephasing dynamics for a combined state of the two spin-$\tfrac{1}{2}$ subsystems \eqref{eq:genstate} one observes that a backflow of information is signified by an increase of the coherences. Of particular importance are the coherences $\kappa_{12}(t)$ \eqref{eq:kappa12} and $\Lambda_{12}(t)$ \eqref{eq:lambda12} as they describe the nonlocal features of the joined state of the two two-level systems. The time evolution of the modulus squared of these coherence factors are connected to the trace distance of the (orthogonal) Bell-states
\begin{align}
|\Psi_I^\pm\rangle&=\frac{1}{\sqrt{2}}(|00\rangle\pm|11\rangle)~,\label{eq:psiI}\\
|\Psi_{II}^\pm\rangle&=\frac{1}{\sqrt{2}}(|01\rangle\pm|10\rangle)~.\label{eq:psiII}
\end{align}
More precisely, the trace distance of these states at time $t$ in the considered model is given by
\begin{align}
\mathcal{D}\bigl(|\Psi_I^+(t)\rangle,|\Psi_{I}^-(t)\rangle\bigr)&=|\kappa_{12}(t)|^2~,\\
\mathcal{D}\bigl(|\Psi_{II}^+(t)\rangle,|\Psi_{II}^-(t)\rangle\bigr)&=|\Lambda_{12}(t)|^2~.
\end{align}

\begin{figure}
    \centering
     \includegraphics[width=0.5\textwidth]{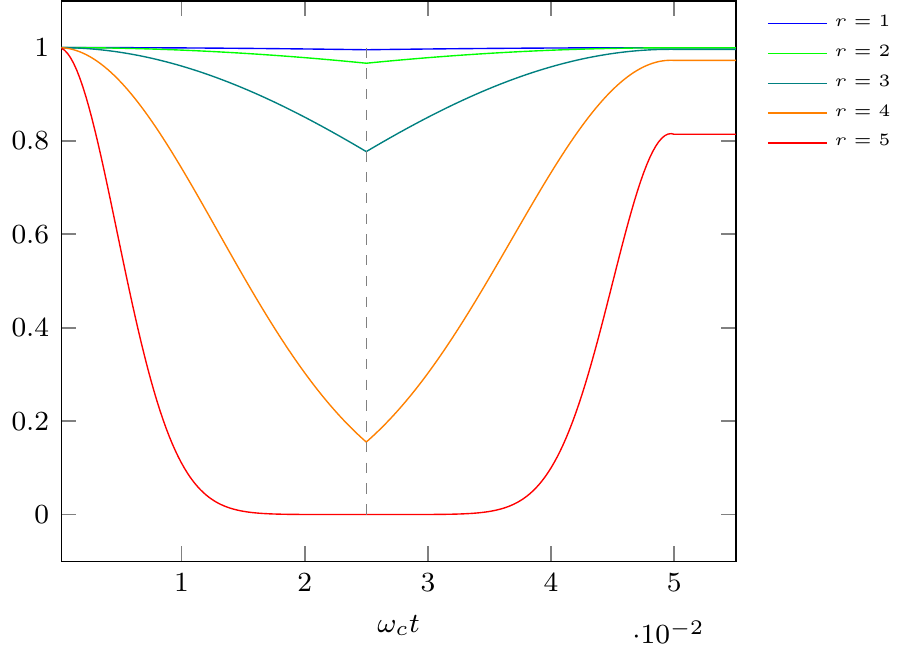}\\
    \caption{Dynamics of $|\Lambda_{12}(t)|^2$ for coupling strength $\alpha_{1,2}=1$ and subsequently applied interactions of length $2.5\cdot10^{-2}$ (in units of $\omega_c$) and several values of the squeezing parameter $r$ for the EPR-state. One obtains for the measure $\mathcal{N}$: $4\cdot10^{-3}$ ($r=1$), $3\cdot10^{-2}$ ($r=2$), $0.2$ ($r=3$), $0.8$ ($r=4$), $0.8$ ($r=5$).} \label{fig:EPRGesamt}
 \end{figure}

Choosing the EPR-state for the two-mode Gaussian state, determined by the covariance matrix \eqref{eq:WernerCov}, one can study the occurence of memory effects for the combined two-level dynamics for subsequently applied interactions of equal length, similar to the analysis done in Ref. \cite{NonlocalTheo}. Performing the maximization included in the measure $\mathcal{N}$ \eqref{eq:measure} numerically, one shows that the maximal increase is given by the orthogonal pair of states $|\Psi_{II}^\pm\rangle$ for positive values of the squeezing. Due to the structure of the coherence factors and the covariance matrix $\boldsymbol{\sigma}_{X,r}^\mathrm{EPR}$ it is clear that changing the sign of the squeezing parameter $r$ transform the functions $|\kappa_{12}(t)|$ and $|\Lambda_{12}(t)|$ into each other.

Fig. \ref{fig:EPRGesamt} shows the dynamics of $|\Lambda_{12}(t)|^2$ for $\alpha_{1,2}=1$ and local interactions of equal length $\omega_c\Delta t=2.5\cdot10^{-2}$ which are in addition turned on and off subsequently. One sees that the rephasing is almost complete for this setup. For $r=4,5$ the non-Markovianity quantified by $\mathcal{N}$ is about $0.8$\,. Hence, there are indeed non-local memory effects in this model which are in addition experimentally accessible. Moreover, going to larger squeezings while reducing the interaction length the effect is amplified yielding full rephasing.

\section{Conclusions}\label{sec:conc}
In this paper we have studied the model introduced in Ref.~\cite{NonlocalTheo} 
with respect to the emergence of non-Markovian effects induced by
nonlocal environmental correlations. We have derived the correct expressions for
the dynamical map of this model for the case of an environmental state which is 
given by a product of correlated two-mode Gaussian states with a covariance 
matrix in standard form. Our results demonstrate that strong nonlocal memory 
effects can be observed if 
one chooses EPR-type Gaussian initial states. Thus, the phenomenon of 
nonlocal memory effects indeed exists in correlated multimode fields.

\acknowledgments \label{sec:acknow}
S.W. thanks the German National Academic Foundation for
support.

\bibliographystyle{mystyleChrono} 

\bibliography{references}

\end{document}